\documentclass[12pt,twoside,fleqn]{article}
\usepackage{espcrc1}
\usepackage{graphicx,epsfig,here}
\begin{document}
\title{Strange Mesons in Dense Nuclear Matter}
\author{Peter Senger
\address{GSI, Planckstr.1,  64291 Darmstadt}
}
\maketitle

\begin{abstract}
Experimental data on the production of kaons and antikaons in
heavy ion collisions at relativistic energies are reviewed with 
respect to in-medium effects. The $K^-/K^+$ ratios measured 
in nucleus-nucleus collisions are 1 - 2 orders of magnitude larger than in
proton-proton collisions.
The azimuthal angle distributions of  $K^+$ mesons    
indicate a repulsive kaon-nucleon potential. 
Microscopic transport calculations consistently explain both the yields and 
the  emission patterns of kaons and antikaons when assuming
that their properties are modified in dense nuclear matter.
The  $K^+$ production excitation functions measured in light and heavy 
collision systems provide evidence for a soft nuclear equation-of-state. 
\end{abstract}

\section{Strangeness in neutron stars and  nuclear fireballs}
Heavy-ion collisions at relativistic energies  provide the unique 
possibility to create a dense and hot nuclear system which can be investigated
experimentally. 
In a reaction between two gold nuclei at a beam energy of 1 AGeV, for 
example, a fireball is produced with baryonic densities up to 3 times 
saturation density and temperatures around 100 MeV. 
Although this transient state exists only for about 10-20 fm/c, it offers
the opportunity to catch a glimpse of nuclear matter far off its ground state.
In nature, similarly extreme conditions exist only in the interior 
of neutron stars. Therefore, experiments at heavy ion accelerators allow to  
address questions which  could not be answered before.
The nuclear equation-of-state at high baryon densities and   
the properties of hadrons in dense nuclear matter are essential for
our understanding of astrophysical phenomena such as the dynamics of a 
supernova and the stability of neutron stars. Moreover,
heavy-ion  experiments  may shed light on the question whether
chiral symmetry is (partly) restored at high baryon densities. 
The study of hadron properties in dense nuclear matter will help to explore a 
fundamental problem  of strong-interaction physics that is the mechanism which 
dynamically breakes chiral symmetry and generates hadron masses.  

\vspace{.5cm}
Strange mesons are regarded as promissing probes both 
for the study of the in-medium properties of hadrons and the 
nuclear equation-of-state \cite{brown91,aichelin}.
Figure 1 displays the range of predictions of various model calculations 
for the total energy of $K$ mesons at rest in nuclear matter as function 
of density (see \cite{schaffner}). 
According to the calculations, the effective mass of a $K^+$ meson
increases moderately with increasing baryon density whereas the effective mass
of $K^-$ mesons decreases significantly. In mean-field calculations,
this effect is caused by a repulsive $K^+$N potential and an attractive 
$K^-$N potential. Microscopic calculations predict a dynamical broadening 
of the $K^-$ meson spectral function which is shifted towards  smaller energies 
\cite{waas,lutz}. The ultimate goal is to relate the
in-medium spectral function of kaons and antikaons to the anticipated
chiral symmetry restoration at high baryon density
 (for details see the contribution of G. Chanfray). 
\begin{figure}[hpt]
\vspace{.cm}
\includegraphics[clip,width=10.cm]{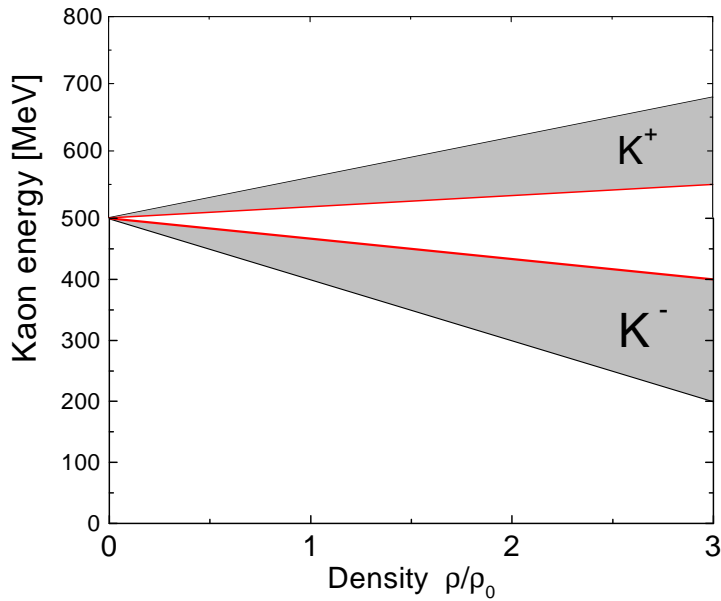}
\vspace{-.5cm}
\caption[]{{\sl
Effective in-medium mass of kaons and antikaons
as function of nuclear density. The grey shaded areas represent the range 
of predictions of various calculations (see \protect\cite{schaffner}).
}
\label{kamed}
}
\end{figure}

H. Bethe and G. Brown pointed out that a strong decrease of the $K^-$ effective
mass with increasing baryon density should have dramatic consequences for 
the stability of neutron stars 
\cite{bro_bet,li_lee_br,hei_jen}. Above a certain value of baryon density 
($\approx$ 3 times saturation density $\rho_0$) 
the total energy of a $K^-$ meson might become  smaller
than the electrochemical potential ($\mu_e\approx$230 MeV).
Then $K^-$ mesons will replace the electrons and form a Bose condensate. 
The condensation of negative bosons enhances the proton to neutron ratio and 
this effect softens the equation-of-state. Consequently, a supernova core
with 1.5 - 2 solar masses will collapse into a black hole rather than form a 
neutron star  \cite{bro_bet}. This mechanism would explain the observation that
the masses of binary pulsars to not exceed significantly the value of
1.5 solar masses.  

\vspace{.5cm}
The predicted behavior of $K$ mesons in dense matter as illustrated in 
figure~\ref{kamed} has not yet been confirmed experimentally. The analysis 
of data on $K^+$-nucleus scattering \cite{friedman1} and kaonic atoms 
\cite{friedman2} found evidence for sizeable 
KN potentials at saturation density and below. Higher densities $-$ which
are of interest for astrophysics $-$ can only
be investigated by heavy-ion experiments.

\vspace{.5cm} 
The challenge is to find experimental observables which can be related 
quantitatively to the searched in-medium effects.    
If the total energy of $K$ mesons is modified in dense nuclear matter 
according to figure~\ref{kamed}, then the energy threshold for 
$K$ meson production will be modified as well. 
$K^+$ and $K^-$ can hardly be produced in 
direct nucleon-nucleon collisions at beam energies below the threshold
which is 1.58 GeV for the reaction $NN\to$$K^+\Lambda$$N$ and 
2.5 GeV for  $NN\to$$K^+K^-$$NN$. Therefore,
kaon production excitation function rises steeply
with increasing beam energy close to and below threshold. This steep rise 
amplifies the effect of a modified in-medium mass on the 
$K$ meson yield.  A moderate enhancement of the  $K^+$ effective mass  of about
10\% will result in an enhanced threshold and, hence, in a 
reduction of the  $K^+$ yield  by about a factor of 2. 
In contrast, the $K^-$ yield will be strongly enhanced as the
effective mass is expected to be reduced substantially 
(see figure~\ref{kamed}). In addition, the 
absorption probability of $K^-$ mesons via the strangeness exchange 
reaction $K^-N\to$$Y\pi$ (with $Y=\Lambda,\Sigma$) will be reduced if the
$K^-$ effective mass decreases.  
Therefore, the yield of kaons and antikaons produced in heavy-ion collisons
at beam energies below threshold is sensitive to 
their effective in-medium masses. Measured cross sections can be 
quantitatively compared to the results of transport calculations which 
take into account also the 
properties of the fireball matter such as density,
compressibility and abundance of excited nucleons.

\vspace{.5cm}
Another measurable effect of in-medium $KN$ potentials is their 
influence on the propagation of kaons and antikaons in heavy-ion collisions.
$K^+$ mesons $-$ which cannot be absorbed as they contain an antistrange quark
$-$ will be repelled from the regions of increased baryonic
density because of the repulsive $K^+N$ potential. In contrast,  
$K^-$ mesons will be attracted \cite{li_ko_br}. Therefore, the 
azimuthal emission pattern of $K$ mesons  is expected to be modified 
according to the in-medium $KN$ potentials and  
the density profile of the nuclear medium.

\vspace{.5cm}
Experiments on kaon and antikaon production and propagation in 
heavy-ion collisions at relativistic energies have been performed with the 
Kaon Spectrometer  \cite{senger} and the FOPI detector \cite{gobbi}
at the heavy-ion synchrotron SIS  at GSI Darmstadt. Results have been 
published in \cite{misko,ritman,ahner,barth,best,shin,laue,crochet}.
The most recent results will be presented in this contribution.
The paper is organized as follows: In section 2 we show that kaons are created 
predominantly at high baryon densities. In section 3 we present evidence 
for in-medium effects on the $K$ meson yields and compare data to results of
microscopic transport calculations. In section 4 we discuss the influence of 
in-medium potentials on the azimuthal emission pattern of $K$ mesons. 
In section 5 we show experimental results   on $K^+$ production which 
indicate that the nuclear equation-of-state is soft.  
Finally, we present a conclusion and an outlook. 

\newpage
\section{Kaons as messengers from the dense nuclear fireball} 
\begin{figure}[hpt]
\vspace{.cm}
  \begin{minipage}[c]{0.5\linewidth}
    \centering
    \mbox{\epsfig{file=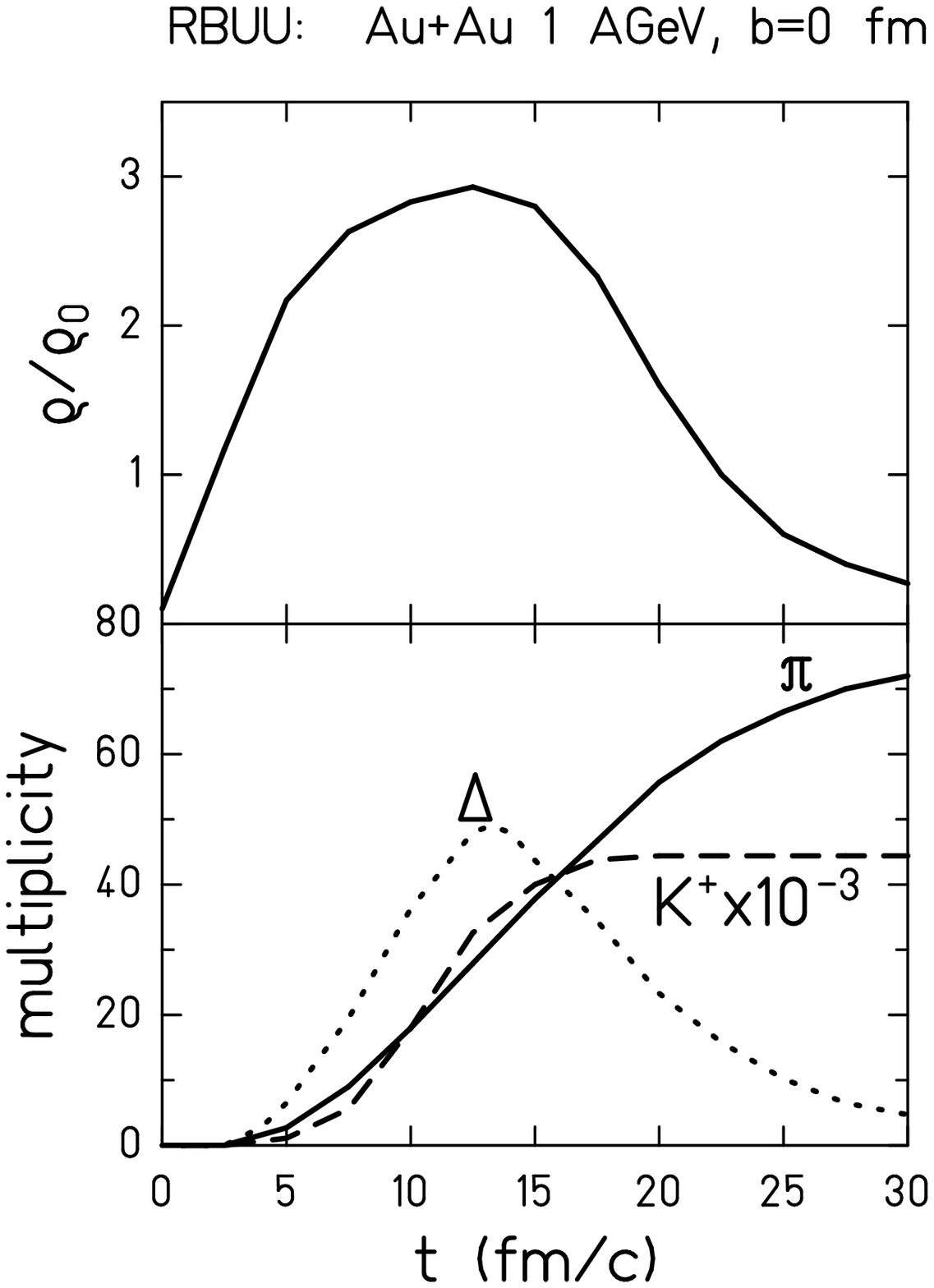,width=11.cm}}
  \end{minipage}\hfill
  \parbox[c]{0.4\linewidth}{
\caption[]{{\sl Upper panel: time evolution of the baryon density 
(nucleons and $\Delta$-resonances) in the fireball of a central
Au+Au collision at a beam energy of 1 A$\cdot$GeV.\\ 
Below: multiplicity of  $\Delta$-resonances (dotted line),
pions  (solid line) and  K$^+$ mesons (dashed line)
as function of time (RBUU-calculation, taken from \protect\cite{fang_ko}).
}}
\label{rbuu_baryon}
}
\end{figure}

In order to obtain information on the high-density phase of a nucleus-nucleus
collision one should measure particles which are created predominantly 
in this phase. This condition is fulfilled for $K^+$ mesons as illustrated     
in figure~\ref{rbuu_baryon} which shows the results of a RBUU calculation 
for a central Au+Au collision at a beam energy of 1 AGeV \cite{fang_ko}.   
The baryon density in the fireball reaches a value of almost three
times saturation density $\rho_0$. The lower part of figure~\ref{rbuu_baryon} 
demonstrates that $K^+$ mesons are produced while the baryon density
exceeds a value of twice $\rho_0$. This is due to $K^+$ production 
via secondary collisions such as $\pi$$N\to$$K^+\Lambda$ or 
$\Delta$$N\to$$K^+\Lambda$$N$
which happen mostly at high baryon densities. 

\begin{figure} [H]
\vspace{-0.cm}
\mbox{\epsfig{file=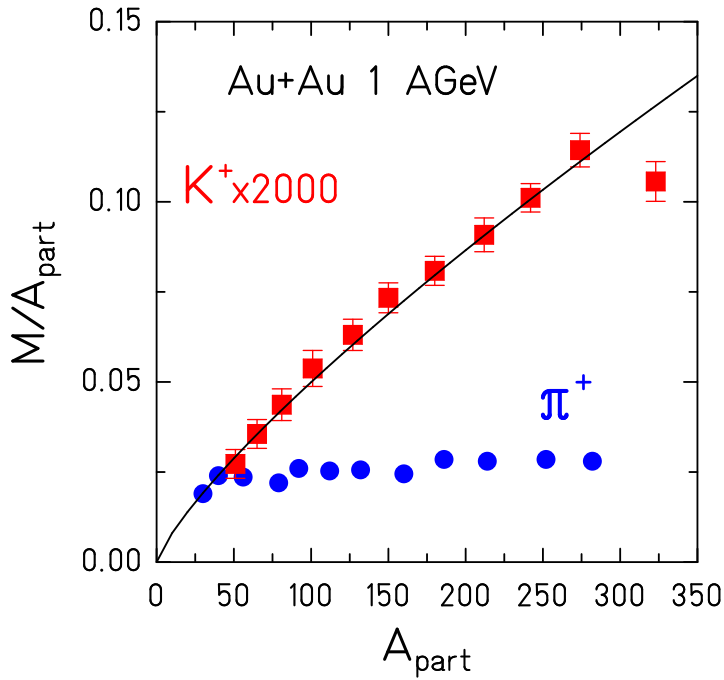,width=13.cm}}
\vspace{-1.cm}
\caption[]{{\sl
K$^+$ and $\pi^+$ multiplicity per participating nucleon
M/A$_{part}$ as a function of A$_{part}$ for Au+Au collisions at 1 AGeV
\protect\cite{mang}. The data are taken
at $\Theta_{lab}$=44$^{\circ}$ and extrapolated to the full solid angle
assuming an isotropic angular distribution in the center-of-mass system.
The line corresponds to a parameterization according to
M$_{K^+}\propto$A$_{part}^{1.8}$.
}
\label{kp_mult_au}
}
\end{figure}

Figure~\ref{rbuu_baryon} cannot be proved directly by experiment as the baryon 
density is not an observable. However, the density dependence of $K^+$ 
production can be studied indirectly by measuring the $K^+$ yield as
function of the collision centrality.  Figure~\ref{kp_mult_au} shows
the $\pi^+$ and $K^+$ multiplicities per number of participating nucleons 
$A_{part}$ as function of $A_{part}$ measured in Au+Au collisions at 1 AGeV 
by the KaoS Collaboration \cite{mang}. In contrast to the pions,
the $K^+$ multiplicity clearly increases more that linearly with $A_{part}$.
This finding indicates  that the nucleons participate more than once in $K^+$ 
production. Such multiple collisions are strongly enhanced at high 
densities. 

\section{In-medium effects on kaon and antikaon yields  }
In this section we review experimental data on $K^+$ and $K^-$ yields as
function of beam energy and  the size of the collision system. We confront 
measured $K^+$ and $K^-$ phase-space distributions with results of transport
model calculations.

\begin{figure} [hpt]
\vspace{0.cm}
\hspace{-.5cm}\mbox{\epsfig{file=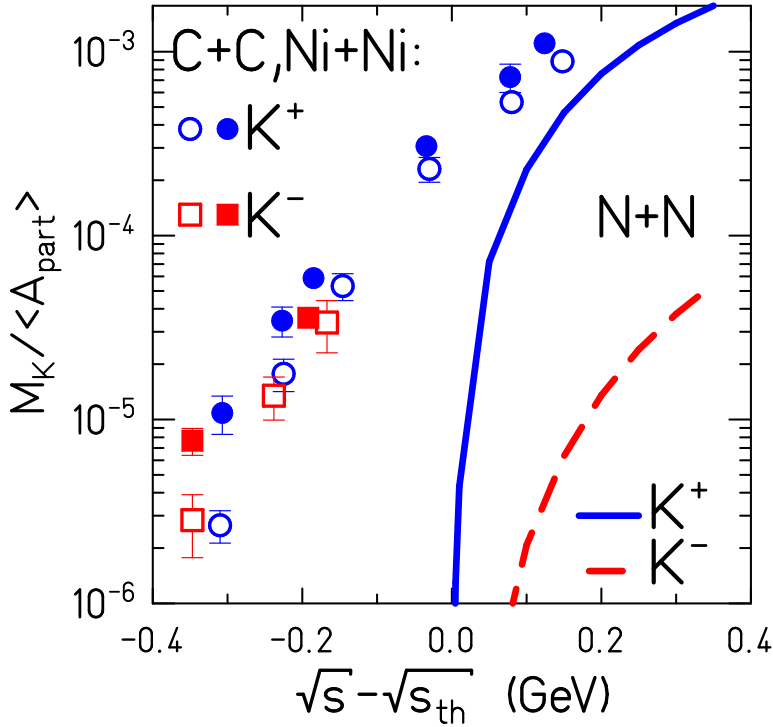,width=14.cm}}
\vspace{-1.cm}
\caption[]{{\sl
$K^+$ (circles) and antikaon (squares) multiplicity per participating nucleon
as a function of the Q-value  for C+C (open symbols) and Ni+Ni (full symbols)
collisions \protect\cite{barth,laue,menzel}.
The lines correspond to parameterizations of the production
cross sections for $K^+$ (solid) and $K^-$ (dashed)
in nucleon-nucleon collisions \protect\cite{sibir95,bra_ca_mo,si_ca_ko}.
}
\label{kp_km_exci}
}
\end{figure}

\vspace{.5cm}
In order to see indications for in-medium effects on $K^+$ and $K^-$
production we compare nucleus-nucleus to proton-proton collisions. 
Fig.~\ref{kp_km_exci} shows the  multiplicity  of $K^+$ and $K^-$ mesons 
per average number of participating nucleons $M_K/<A_{part}>$
as function of the Q-value in the $NN$ system. 
The data were measured in C+C and Ni+Ni collisions by
the KaoS Collaboration \cite{barth,laue,menzel}. The Q-value is defined as the 
energy above threshold. The lines represent parameterizations
of the available proton-proton data
averaged over the isospin channels \cite{sibir95,bra_ca_mo,si_ca_ko}.
In ''nucleon-nucleon'' collisions the kaon multiplicity exceeds the 
antikaon multiplicity by 1-2 orders of magnitudes at the same Q-value. 
This large difference 
has disappeared for nucleus-nucleus collisions where the kaon and antikaon 
data nearly fall on the same curve.
This observation indicates that the production of antikaons is 
strongly enhanced in nucleus-nucleus collisions.

\vspace{.5cm}

\begin{figure}[H]
\vspace{-.0cm}
\begin{minipage}[c]{0.6\linewidth}
    \centering
\mbox{\epsfig{file=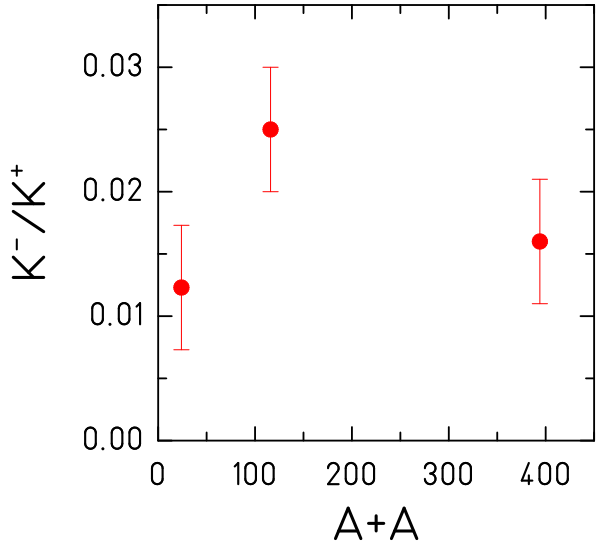,width=11.cm}}
 \end{minipage}\hfill
  \parbox[c]{0.35\linewidth}{
\caption{{\sl $K^-/K^+$ ratio measured in C+C, Ni+Ni and Au+Au collisions
at a beam energy of 1.5 AGeV (preliminary)
\protect\cite{laue,menzel,foerster}. 
}}
\label{karat}
}
\end{figure}

\vspace{.5cm}
In the following we discuss the possibility to study in-medium effects 
in $K$ meson production by varying the size of the collision system. 
Figure~\ref{karat} presents the $K^-/K^+$ ratio measured in 
C+C, Ni+Ni and Au+Au collisions at a beam energy of 1.5 AGeV
\cite{laue,menzel,foerster}. The ratio is 
approximately constant for the three systems, although the reabsorption
probability for $K^-$ mesons should be very different:      
The $K^-$ mean free path is about 1.5 fm  at saturation density.
In a  geometrical model, the losses of K$^-$ mesons due
to strangeness exchange $K^-N\to$$Y\pi$ are estimated to be more than
10 times larger in Au than in C nuclei.
Therefore, the data in figure ~\ref{karat} suggest that in the 
Au+Au system the absorption of $K^-$ is compensated by an enhanced 
production and/or that the $K^-$ absorption process is suppressed.
Both effects can be caused by a reduced in-medium effective mass of antikaons.

\begin{figure}[hpt]
\begin{center}
\vspace{-0.cm}
\mbox{\epsfig{file=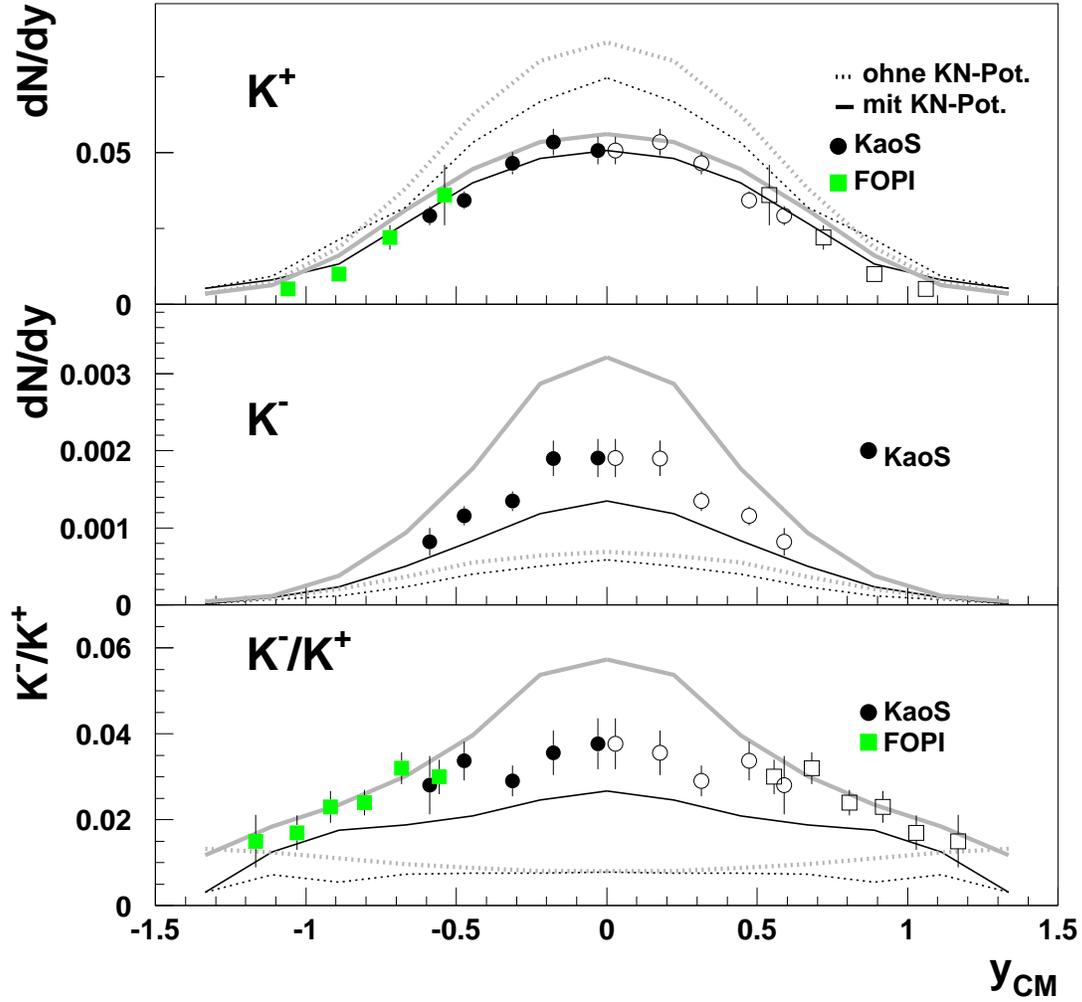,width=16.cm}}
\vspace{-.cm}
\caption{{\sl Multiplicity density distributions of $K^+$ (upper panel)
and $K^-$ mesons (center panel)
for near-central (b $<$ 4.4 fm) Ni+Ni collisions at 1.93 AGeV.
Circles: KaoS data \protect\cite{menzel}, squares: 
FOPI data \protect\cite{best,wisnie}. 
The measured data (full symbols) are mirrored at y$_{CM}$=0 (open symbols).
Lower panel: K$^-$/K$^+$ ratio.
The data are compared to BUU transport calculations
(black lines \protect\cite{cass_brat}, grey lines \protect\cite{li_brown}). 
Solid lines: with in-medium effects. Dotted lines:
without in-medium effects. 
}}
\label{dndy}
\end{center}
\end{figure}

Now we compare measured phase-space distributions of 
$K^-$ and $K^+$ mesons to the results of transport codes in order to 
analyze more quantitatively the in-medium effects.   
Figure~\ref{dndy} shows the $K^+$ and $K^-$ multiplicity densities dN/dy
and their ratio for near-central 
Ni+Ni collisions at 1.93 AGeV as function of the rapidity y$_{CM}$.
The rapidity is defined here as $y_{CM} = y - 0.5\times y_{proj}$.
The value $y_{CM} = 0$  corresponds to
midrapidity and $y_{CM} = \pm 0.89$ to projectile or target rapidity,
respectively.
Figure~\ref{dndy} combines data measured by the KaoS Collaboration (circles) 
\cite{menzel} and by the FOPI Collaboration (squares) \cite{best,wisnie}. 
The KaoS data were analyzed for
the most central 620$\pm$30 mb of the reaction cross section (corresponding
to  impact parameters smaller than b = 4.4 fm in a sharp cutoff model)
whereas the FOPI data were analysed for the most central 350 mb (b = 3.3 fm).
Therefore, the FOPI multiplicities  should be  higher by about 20\%.
Nevertheless, the $K^+$ multiplicity densities and the $K^-/K^+$ ratios 
measured by the two experiments in the overlaping rapidity range agree 
with each other within the experimental errors.  

\vspace{.5cm}

\vspace{.5cm}
The data are compared to results of RBUU transport calculations  
performed by Li and Brown (grey lines  \cite{li_brown}) 
and Cassing and Bratkovskaya (black lines  \cite{cass_brat}).
The dashed lines represent the results of  calculations
with bare kaon and antikaon masses whereas the solid lines are calculated with
in-medium masses. 
The ''bare mass'' calculations clearly overestimate the $K^+$ yield
and underestimate the $K^-$ yield. The results of the ''in-medium'' 
calculations reproduce well the $K^+$ data  but not the $K^-$ data
which are overestimated by Li and Brown and underestimated by Cassing and
Bratkovskaya. The data now allow for a fine tuning of the strength of the 
$K^-N$ potential which is a free parameter in those models. 
\section{In-medium effects on the emission pattern of $K^+$ mesons}
In this section we discuss the influence of the in-medium $KN$ potentials on 
the propagation of $K^+$ mesons inside the nuclear fireball. 
The $K^+$ mesons are expected to be repelled 
from the bulk of nucleons due to the repulsive $K^+N$ potential. 
The appropriate observable for this effect is the azimuthal emission pattern 
which reflects the collision geometry.  

\begin{figure}[hpt]
\vspace{-1.5cm}
\begin{center}
\hspace{0.cm}\epsfig{file=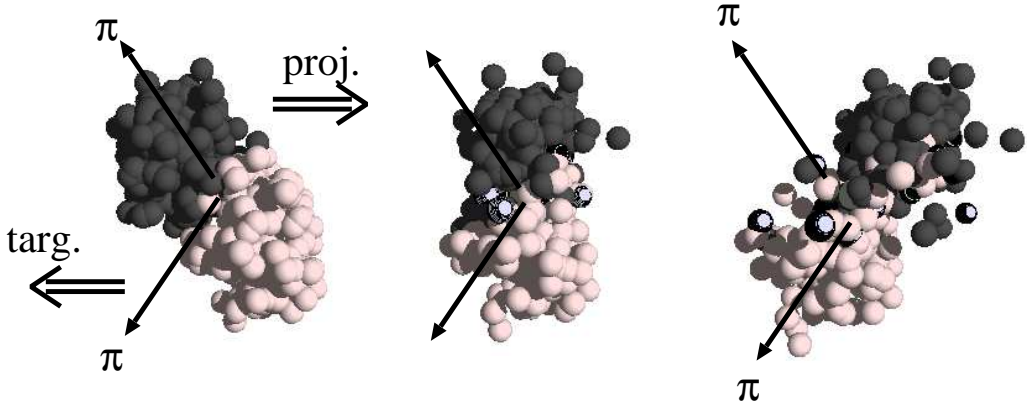,width=14cm}
\vspace{-.5cm}
\caption{{\sl Sketch of an Au+Au collision at 1 AGeV with an impact parameter
of 7 fm as calculated by a QMD tranport code \protect\cite{hartnack}.
The snapshots are taken at 6.5 fm/c (left),  12.5 fm/c (middle) and  18.5 fm/c
(right).  The arrows indicate the position of the spectrometer at target
rapidity. 
}}
\label{model}
\end{center}
\vspace{-.5cm}
\end{figure}

\vspace{.5cm}
Figure~\ref{model} sketches  the nuclear matter distribution in 
a semi-central Au+Au collision at 1 AGeV  for the time steps 6.5 fm/c, 
12.5 fm/c and  18.5 fm/c as calculated by a QMD transport 
code \cite{hartnack}. The trajectories of target and projectile define
the reaction plane. This collision system was studied by the KaoS Collaboration 
which measured the azimuthal angular distributions of protons, pions and kaons.
The particles were detected at a laboratory angle of $\Theta_{lab}=84^o$
corresponding to a rapidity interval of $0 \le y/y_{proj} \le 0.2$.
The direction to the spectrometer is indicated in figure~\ref{model} 
by the arrows.

\vspace{.5cm}
\begin{figure}[H]
\vspace{-.5cm}
\mbox{\epsfig{file=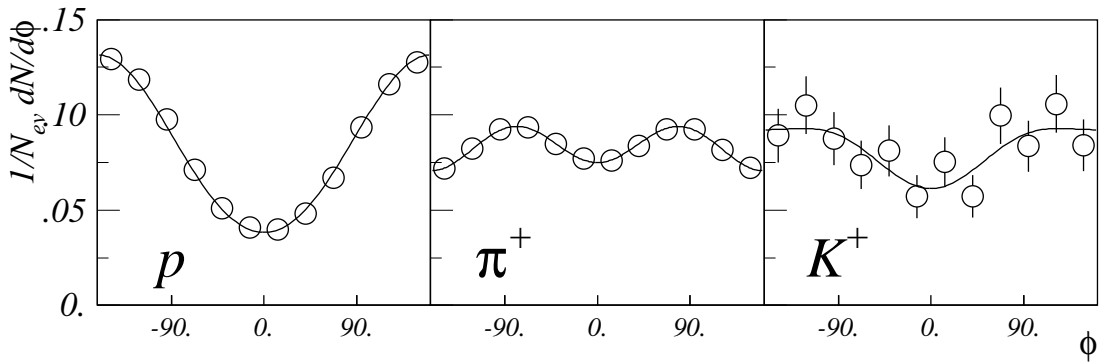,width=15.cm}}
\vspace{-.5cm}
\caption{{\sl Azimuthal angle distribution of protons (left), 
pions (center) and $K^+$ mesons (right) measured 
in Au+Au collisions at 1 AGeV around target rapidity by the 
KaoS Collaboration \protect\cite{inga}. 
The accepted range in transverse momentum is
0.25 GeV/c$ < p_t <$ 0.75 GeV/c. The solid line represents a Fourier expansion
fitted to the data (see text)     
}}
\label{kaflow}
\end{figure}

The azimuthal emission pattern of protons, pions and $K^+$ mesons
as measured in semi-central Au+Au collisions at 1 AGeV are shown in 
figure~\ref{kaflow} \cite{inga}. 
The data are analyzed for impact parameters of $b>5$ fm 
and transverse momenta of 0.25 GeV/c$ < p_t <$ 0.75 GeV/c. 
The reaction plane is defined by the azimuthal angles 
$\phi$ = 0$^{\circ}$,180$^{\circ}$ together with the beam axis. 
A dip in the angular distribution  around $\phi$ = 0$^{\circ}$
corresponds to a predominant in-plane emission (socalled sideward flow) 
whereas the bumps at $\phi = \pm90^{\circ}$  reflect an enhanced
emission perpendicular to the reaction plane.
The solid lines represent the Fourier expansion 
\begin{center}
$dN/d\phi \propto [1 + 2v_1cos(\phi) + 2v_2cos(2\phi)]$ 
\end{center}
which is fitted to the measured azimuthal distributions. The Fourier 
coefficients 

$v_1 = <cos(\phi)>$  and $v_2 = <cos(2\phi)>$ characterize the strength of 
the emission sidewards (in-plane) and out-of-plane, respectively.

\vspace{.5cm}
The proton distributions exhibit a very pronounced sideward flow signal.
The pions are preferentially emitted perpendicular to the reaction plane. 
This effect is caused by the spectator fragments which shadow the pions
emitted in-plane (see figure~\ref{model}). 
The preferential in-plane emission of $K^+$ mesons 
indicates that the bulk of the $K^+$ mesons is emitted early, before the
target spectator is able to shadow the kaons. From figure~\ref{model} one 
can estimate that the shadowing by the target spectator sets in after
approximately 15 fm/c. This result is in remarkable agreement with 
the time evolution of kaon production as calculated by transport models 
(see figure~\ref{rbuu_baryon}). The kaon flow signal
is even more pronounced than the flow signal of the high-energy pions
which are not shown in figure~\ref{kaflow} \cite{wagner1}. 
This effect  indicates that $K^+$ mesons are strongly repelled from the 
participant and spectator matter.

\begin{figure}[hpt]
\mbox{\epsfig{file=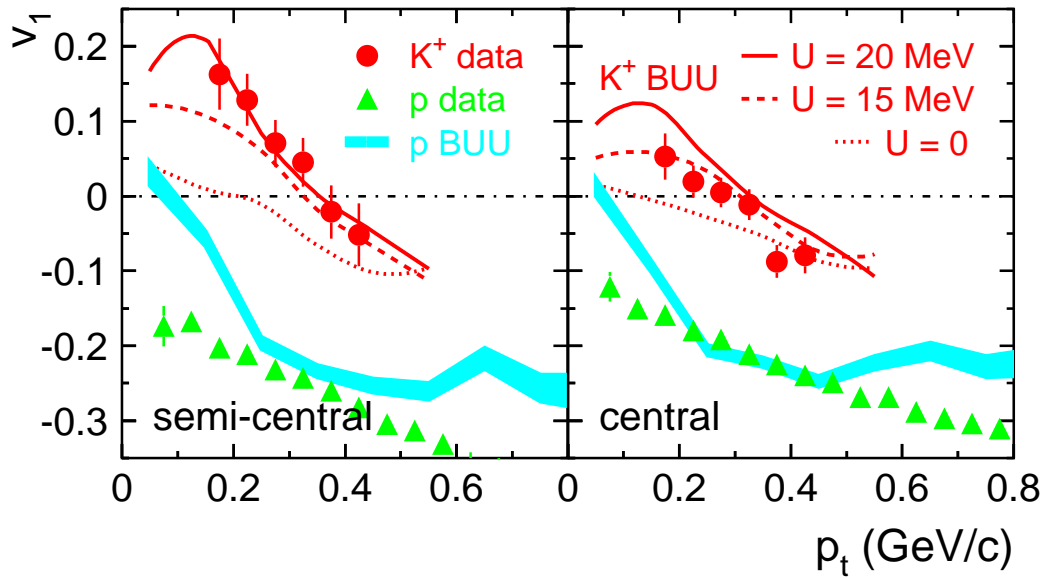,width=14.cm}}
\vspace{-1.cm}
\caption{{\sl
Flow coefficient $v_1$ for protons (triangles) and $K^+$ mesons (dots)
versus transverse momentum $p_t$
measured in semi-central (left) and central (right) Ru+Ru collisions 
at 1.69 AGeV by the FOPI Collaboration. The detector covers the
rapidity range of $-1.2 < y^0 < -0.65$. The grey shadow illustrate
protons from BUU calculations, the lines represent $K^+$ mesons
calculated for different $K^+N$ potentials as indicated. 
Taken from \protect\cite{crochet}. 
}}
\label{kaflow_fopi}
\end{figure}

The FOPI Collaboration has studied the directed flow of kaons in more detail. 
Figure ~\ref{kaflow_fopi} shows the $v_1$ coefficient for protons 
(triangles) and $K^+$ mesons (dots) as function of transverse momentum  
measured in semi-central (left) and central (right) Ru+Ru collisions
at 1.69 AGeV \cite{crochet}.   
The data are taken around target rapidities. It can be seen that in 
semicentral collisions the sign of $v_1$ is opposite for protons and 
low-momentum $K^+$ mesons, i.e the kaons exhibit ''antiflow''  below
transverse momenta of about $p_t$=0.35 GeV/c. 
This observation is explained within 
transport calculations by a repulsive $K^+N$ potential of U = 20 MeV 
at saturation density (solid lines). In central collisions, the 
low-momentum $K^+$ mesons are emitted more or less isotropically and thus
a weaker $K^+$N potential is required  by the calculations in order to describe
the data. It should be noted that the calculations fail to describe 
the proton data. This discrepancy and the simplified assumption 
of a momentum-independent in-medium potential deserve further 
theoretical efforts.

\begin{figure}[hpt]
\vspace{-0.cm}
    \centerline{
\hspace{-0.cm}\mbox{\epsfig{file=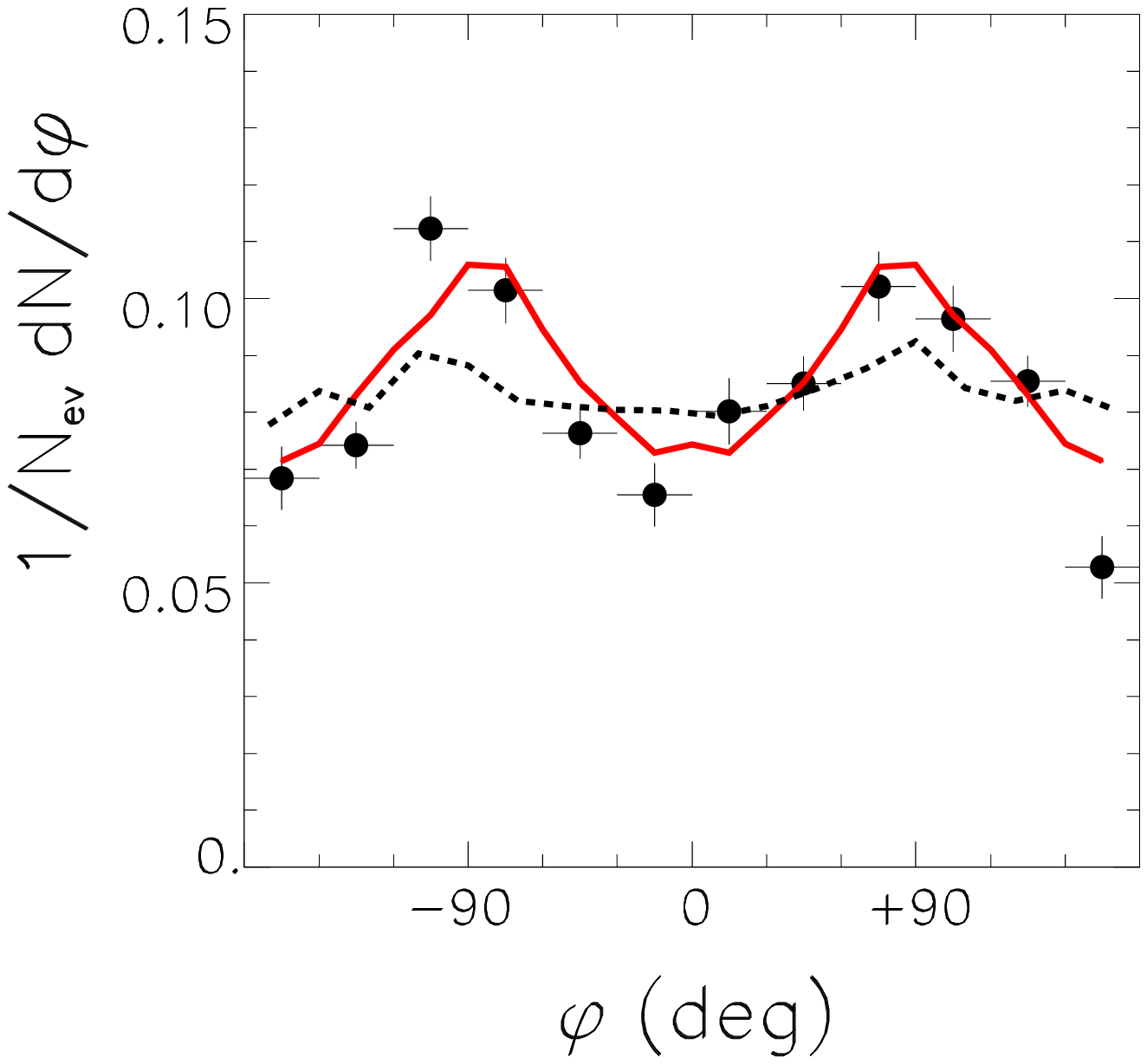,width=6.cm}}
\hspace{.5cm}\mbox{\epsfig{file=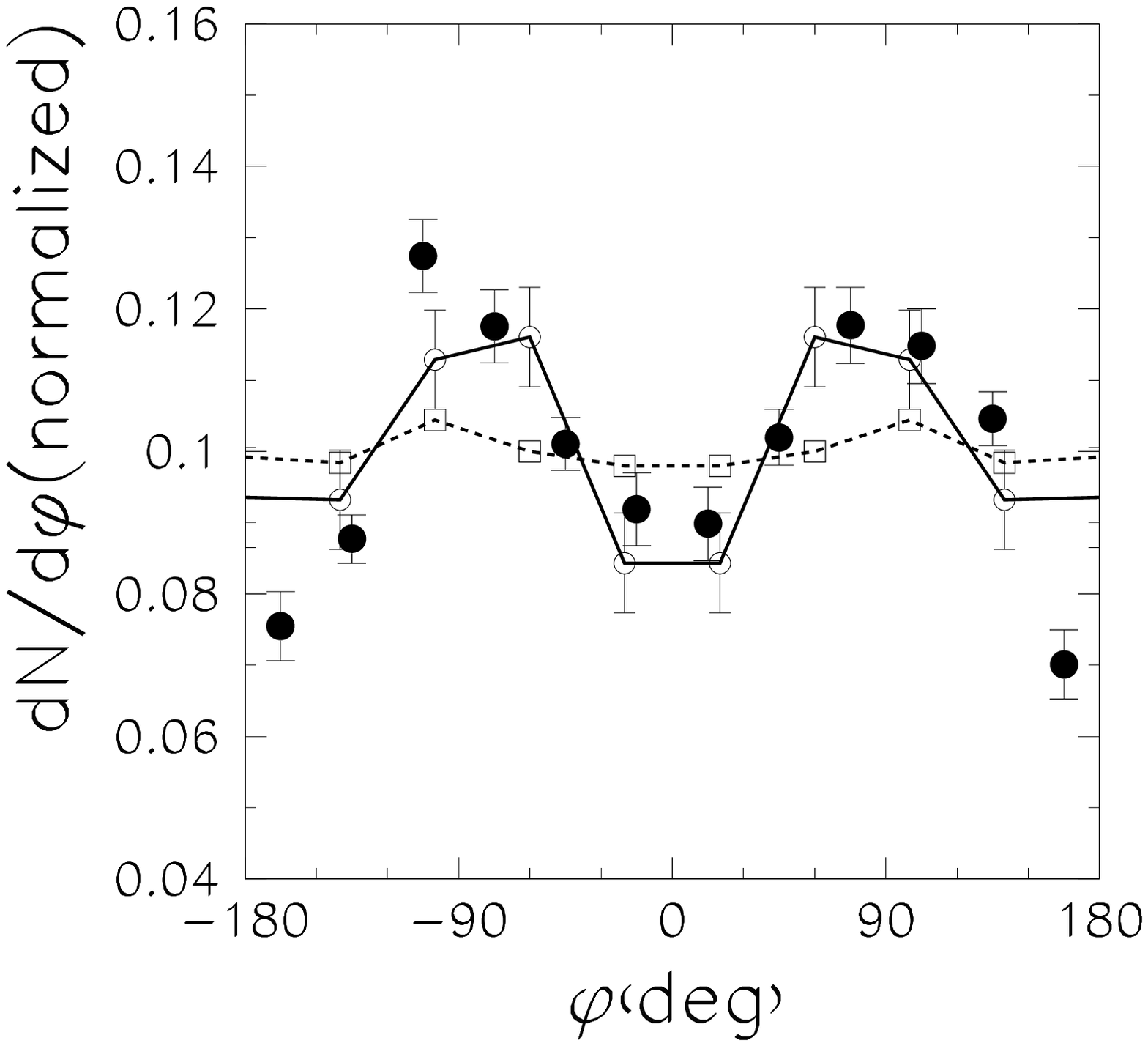,width=6.5cm}}
}
\vspace{-.5cm}
\caption[]{{\sl 
$K^+$ azimuthal distribution for semi-central Au+Au collisions at 1 AGeV
(full dots). The data are analyzed at $0.4< y/y_{proj}< 0.6$ (left)
and $0.2< y/y_{proj}< 0.8$ (right) \protect\cite{shin}.
The lines represent results of transport calculations from the
Stony Brook group using a RBUU model
(left \protect\cite{li_ko_br}) and a
QMD model from the T\"ubingen group (right \protect\cite{wang99}).
Both models take into account
rescattering, the QMD version also considers Coulomb effects.
Solid lines: with in-medium $KN$ potential.
Dashed lines: without  in-medium $KN$ potential.
}}
\label{shin}
\end{figure}

\vspace{.5cm}
The KaoS Collaboration has measured the azimuthal emission pattern of $K^+$ 
mesons  in Au+Au collisions at 1 AGeV  around midrapidity \cite{shin}.   
Fig.~\ref{shin} presents the dN/d$\phi$ distribution of  
kaons which were accepted within a range
of transverse momenta of 0.2 GeV/c $\leq$ p$_t$$\leq$ 0.8 GeV/c
for two ranges of normalized rapidities
0.4 $\leq$ y/y$_{proj}$ $\leq$ 0.6 (left) and
0.2 $\leq$ y/y$_{proj}$ $\leq$ 0.8 (right)
with y$_{proj}$ the projectile rapidity.
The data are corrected for the uncertainty of the
determination of the reaction plane on the basis of a Monte Carlo simulation.

\vspace{.5cm}
The $K^+$ emission pattern clearly is peaked at $\phi$=$\pm$90$^0$
which is perpendicular to the reaction plane. Such a behavior is known from
pions \cite{brill} which interact with the spectator fragments.
In the case of $K^+$ mesons, however, the anisotropy can be explained  
by transport calculations only if  
a repulsive in-medium $K^+N$ potential is assumed. This is demonstrated 
by the solid lines in fig.~\ref{shin}. A flat distribution (dotted lines) is
expected when neglecting the in-medium potential and only 
taking into account $K^+$ rescattering (\cite{li_ko_br}, left) 
and additional Coulomb effects (\cite{wang99}, right). 

\vspace{.5cm}
In summary, the $K^+$ azimuthal emission pattern exhibits structures 
which contradict the naive picture of a  long mean free path in 
dense nuclear matter. The particular features  of sideward flow and the 
pronounced out-of-plane emission around midrapidity
indicate that  $K^+$ mesons are repelled from
the nucleons as expected for a repulsive $K^+N$ potential. 
The key observable for the study of $KN$ potentials is the azimuthal 
emission pattern of $K^-$ mesons. If an attractive 
$K^-N$ potential exists, the $K^-$ mesons will be emitted almost 
isotropically in semi-central Au+Au collisions. This is in contrast 
to what one expects for a particle being absorbed  in nuclear matter
which is flowing nonisotropically \cite{wang99}.  Such an experiment has been 
performed by the KaoS Collaboration. The analysis is in progress.    

\section{K$^+$ production and the nuclear equation-of-state} 
Microscopic transport calculations claim that the yield  of kaons created
in collisions between heavy nuclei at subthreshold beam energies
(E$_{beam}$ = 1.58 GeV for $NN\to$$K^+\Lambda$$N$)
is sensitive to  the compressibility of
nuclear matter at high baryon densities \cite{aich_ko,li_ko}.
This sensitivity is due to the production mechanism of $K^+$ mesons.
At subthreshold  beam energies,
the production of kaons requires multiple nucleon-nucleon collisions
or secondary collisions  such as $\pi$$N\to$$K^+\Lambda$ and 
$\Delta$$N\to$$K^+\Lambda$$N$. These processes
are expected to occur predominantly at high baryon densities, and
the densities reached in the fireball depend on the nuclear equation-of-state
\cite{fuchs97}.

\begin{figure}[hpt]
\vspace{-0.cm}
\mbox{\epsfig{file=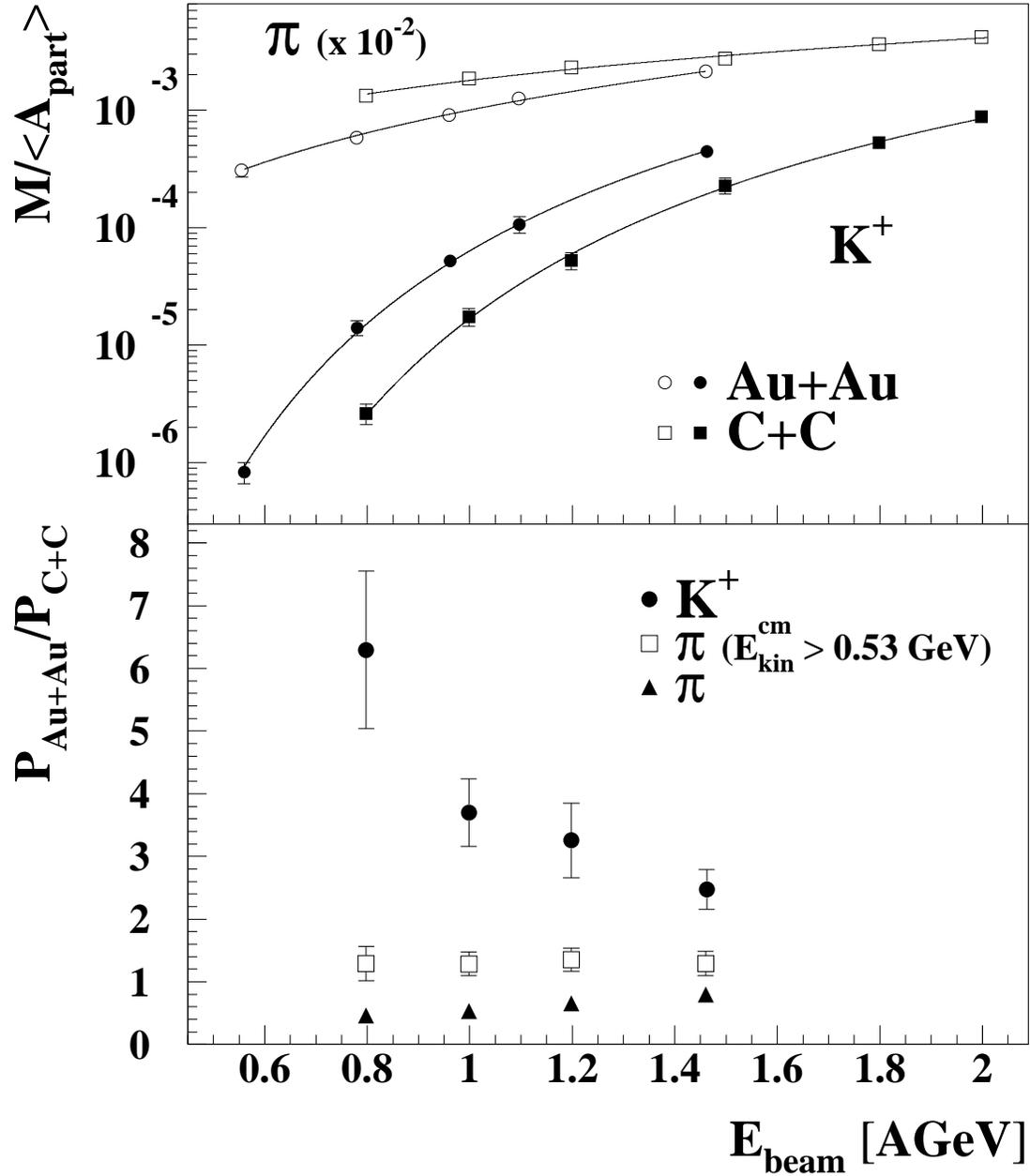,width=16.cm}}
\vspace{-1.cm}
\caption{{\sl
Upper panel:
Pion and $K^+$ multiplicity per participating nucleon $P~=~M/<A_{part}>$
for Au+Au and C+C collisions as function of the projectile energy per nucleon.
The pion data include charged and neutral pions (see text). The lines
represent a fit to the data.
Lower panel:
Ratio of the multiplicities per participant  (Au+Au over C+C collisions)
for $K^+$ mesons (full circles), pions (full triangles) and high-energy
pions (E$^{cm}_{kin}>$ 530 MeV, open squares) as function of the projectile
energy per nucleon. Taken from \protect\cite{sturm}.
}}
\label{pikapexi}
\end{figure}

$K^+$ mesons are well suited to probe the properties of the
dense nuclear medium because of the absence of  absorption.
They  contain an antistrange quark and hence emerge as
messengers from  the dense phase of the collision.
In contrast, the pions created in the high density phase of the
collision are likely to be reabsorbed and most of them
will leave  the reaction zone in the late phase \cite{bass}.

\vspace{.5cm}
However, the $K^+$ yield does not only depend on the nuclear compressibility
but also on the in-medium $K^+N$ potential, as we have seen in the preceding
sections. The KaoS Collaboration pursued the idea to disentangle 
these two effects by studying $K^+$ production in a very light 
($^{12}$C+$^{12}$C) and a heavy collision system ($^{197}$Au+$^{197}$Au)
at different beam energies near threshold \cite{sturm}.
In the heavy Au+Au system the average baryonic density - achieved
by the pile-up of nucleons - is significantly higher than 
in  C+C collisions \cite{cass_brat}.
Moreover, the maximum baryonic density reached in Au+Au collisions
depends on the nuclear compressibility \cite{aichelin,li_ko}
whereas in the small C+C system this dependence is very weak \cite{fuchs00}.
Due to the fact that kaons are produced in multiple steps the  kaon yield 
depends at least quadratically on the density. On the other hand, 
the repulsive $K^+N$ potential is assumed to depend nearly
(or less than) linearly on the baryonic density \cite{schaffner} and thus
reduces the kaon yield accordingly.
Therefore, the influence of the nuclear compressibility on the $K^+$ yield 
in Au+Au collisions should be measurable,
and this effect is expected to  increase with decreasing beam energy. 

\begin{figure}[hpt]
\vspace{-.5cm}
\begin{minipage}[c]{0.6\linewidth}
    \centering
\mbox{\epsfig{file=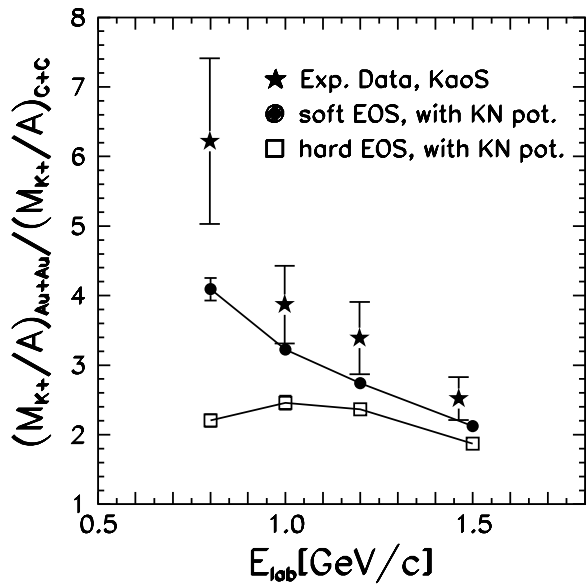,width=8.cm}}
 \end{minipage}\hfill
  \parbox[c]{0.37\linewidth}{
\caption{{\sl $K^+$ multiplicity ratio $P_{Au+Au}/P_{C+C}$ as function
of the beam energy for inclusive reactions. 
The data (stars) are compared to results of  
QMD transport model calculations  based on a soft (full dots) and a hard
(open squares) nuclear equation-of-state \cite{fuchs00}.
}}
\label{kaprat}
}
\end{figure}
The multiplicity of K$^+$ mesons has been measured in inclusive C+C collisions 
at beam energies between 0.8 and 2 AGeV \cite{laue} 
and in inclusive  Au+Au collisions between  0.6 and 1.5 AGeV \cite{sturm}. 
The upper panel of figure~\ref{pikapexi} shows the inclusive pion and 
$K^+$ multiplicity 
per average number of participating nucleons $M/<A_{part}>$
for C+C and Au+Au collisions as a function of beam energy. The average number 
of participating nucleons is given by $<A_{part}>$=A/2 according to a 
geometrical model. 
The lower panel of figure~\ref{pikapexi} presents the ratios
of the particle multiplicities  measured in Au+Au and C+C collisions
$P_{Au+Au}/P_{C+C}$ with $P = M/<A_{part}>$.  
The pion ratio (triangles) is well below unity which indicates
pion absorption in the heavy system.
In contrast, the  $K^+$ mesons
ratio $P_{Au+Au}/P_{C+C}$ (full dots) is above two and still increases with 
decreasing beam energy. 

\vspace{.5cm}
Recently, the $K^+$ ratio  $P_{Au+Au}/P_{C+C}$ has been calculated 
also  for inclusive reactions with 
QMD transport models \cite{fuchs00,hart_aich}. Figure~\ref{kaprat} shows the 
results of calculations \cite{fuchs00} 
which are performed with two values for the 
compression modulus:  $\kappa$ = 200 MeV (a ''soft'' equation-of-state, 
full dots) and $\kappa$ = 380 MeV (a ''hard'' equation-of-state).
These calculations take into account a repulsive kaon-nucleon potential
and use momentum-dependent Skyrme forces.
Figure~\ref{kaprat} clearly demonstrates that the calculation based on a 
soft equation-of-state reproduces the trend of the experimental data (stars).
The ratio $P_{Au+Au}/P_{C+C}$ has the advantage that many uncertainties 
of both  the model calculation and the experiment cancel in this
presentation. Therefore, it is a promissing task to improve the model inputs 
in order to achieve better agreement with the data. 
\section{Conclusions and outlook }
We have presented data on the production and propagation of strange mesons
in heavy-ion collisions in the  SIS energy range. 
The $K^-/K^+$ ratio as function of the Q-value was found to be a factor
of 10 - 100 larger in nucleus-nucleus that in proton-proton collisions.
The $K^-/K^+$ ratio is almost independent of the size of the collision system
although the absorption probability of $K^-$ mesons should dramatically 
increase with increasing system size.
These observations indicate that the yield of $K^-$ mesons is enhanced and/or
their absorption is reduced in dense nuclear matter. 
According to transport model calculations, the enhanced yield of $K^-$ mesons
is caused by their reduced effective mass in the medium. 
In Au+Au collisions at 1 AGeV, the $K^+$ mesons are emitted preferentially
into the reaction plane at target rapidities and out-of-plane at midrapidity.
In Ru+Ru collisions at 1.69 AGeV, the low momentum $K^+$ mesons are 
preferentially emitted opposite to the nucleons. 
These effects are explained within transport calculations by a repulsive 
$K^+N$ potential. 

The measurement of the $K^-$ azimuthal emission pattern
is considered to be a key experiment in this respect: 
if a strongly attractive $K^-N$ potential exists, $K^-$ mesons
will be emitted almost isotropically although the nucleons
are flowing nonisotropically in  a semicentral nucleus-nucleus collision.    
More information on the in-medium properties of
kaons and antikaons can be expected 
from production yields in proton-nucleus collisions. 
Both experiments have been performed by the  KaoS Collaboration, 
and their analysis is in progress. 

In the fall of this year the new Dilepton Spectrometer HADES at SIS/GSI
will take first data. This will open a new era of studies of vector 
mesons and hadron  properties in dense nuclear matter.  
\section{Acknowledgement}
I would like to thank J. Aichelin, E. Bratkowskaya, W. Cassing, C. Fuchs  and 
M. Lutz for valuable discussions. I thank the  members of the FOPI Collaboration
P. Crochet, A. Devismes, R. Kutsche and W. Reisdorf for supplying 
me with their data. It is a pleasure to thank
my colleagues from the KaoS Collaboration  who measured and analyzed 
most of the data presented in this article:
I. B\"ottcher, A. F\"orster, E.Grosse, P. Kocz\'on,  B. Kohlmeyer, M. Menzel, 
L.Naumann, H. Oeschler, E. Schwab, W.Scheinast, Y. Shin, H. Str\"obele,   
C. Sturm,  F. Uhlig,  A.Wagner and W.Walu\'s.

\end{document}